\documentclass[5p,times,merge,number,sort&compress]{elsarticle}

\usepackage[T1]{fontenc}
\usepackage[utf8]{inputenc}
\usepackage[english]{babel}

\usepackage{graphicx}
\usepackage{isotope}
\usepackage{physics}
\usepackage{xcolor}
\usepackage{subcaption}
\usepackage{hyperref}
\usepackage{amsmath}
\usepackage{multirow}
\usepackage{wasysym}
\usepackage{fdsymbol}

\DeclareMathAlphabet{\curly}{OMS}{cmsy}{m}{n}

\newcommand*{\elem}[2]{\ensuremath{\isotope[#2]{\mathrm{#1}}}}
\newcommand*{\lelem}[2]{\ensuremath{\isotope[#2\,][\Lambda]{\mathrm{#1}}}}

\definecolor{FGOrange}{rgb}{1.0, 0.5490196078431373, 0.0}
\definecolor{FGgreen}{rgb}{.0, 0.5, 0.0}
\definecolor{FGred}{rgb}{1.0, 0.0, 0.0}

\definecolor{myblue}{rgb}{0.03137254901960784, 0.11372549019607843, 0.34509803921568627}
\definecolor{mygreen}{rgb}{0.0, 0.26666666666666666, 0.10588235294117647}
\definecolor{mypurple}{rgb}{0.5019607843137255, 0.0, 0.14901960784313725}

\begin{document}

\title{Hyperon-Nucleon Interaction Constrained by Light Hypernuclei}

\address[tud]{Institut f\"ur Kernphysik, Fachbereich Physik, Technische Universit\"at Darmstadt, Schlossgartenstr. 2, 64289 Darmstadt, Germany}
\address[hfhf]{Helmholtz Forschungsakademie Hessen f\"ur FAIR, GSI Helmholtzzentrum, 64289 Darmstadt, Germany}

\author[tud]{Marco~Kn\"oll}
\ead{mknoell@theorie.ikp.physik.tu-darmstadt.de}
\author[tud,hfhf]{Robert~Roth}
\ead{robert.roth@physik.tu-darmstadt.de}

\date{\today}

\begin{abstract}

\noindent \textit{Ab initio} structure calculations for p-shell hypernuclei have recently become accessible through extensions of nuclear many-body methods, such as the no-core shell model, in combination with hyperon-nucleon interactions from chiral effective field theory. 
However, the low-energy constants in these hyperon-nucleon interactions are poorly constraint due to the very limited amount of experimental scattering data available.
We present a hyperon-nucleon interaction that is additionally constrained by experimental ground-state and spectroscopic data for selected p-shell hypernuclei and, thus, optimized for hypernuclear structure calculations. 
We show that the previous overestimation of the hyperon separation energies in the p-shell is remedied and discuss the significantly improved description of the \lelem{He}{} isotopic chain.
We further discuss the uncertainty quantification for hypernuclear observables on the many-body level, obtained through a novel machine-learning tool.

\end{abstract}

\maketitle

\paragraph{Introduction}

In recent years, the endeavours to understand the fundamental aspects of the strong interaction have extended beyond purely nucleonic approaches to a more general baryonic picture, raising questions such as the description of many-body systems with strangeness or the hyperon puzzle in neutron stars \cite{Bombaci2017hyperon}.
In the regime of s and p-shell hypernuclei, \textit{ab initio} structure calculations have become accessible through extensions of nuclear many-body methods, such as Faddeev-Yakubovsky calculations \cite{Nogga2002hypernuclei,Nogga2014light}, Gaussian expansion methods \cite{Hiyama2009structure}, Quantum Monte-Carlo approaches \cite{Lonardoni2013effects,Lonardoni2013auxiliary} and, more recently, nuclear lattice calculations \cite{Frame2020impurity} and no-core shell model (NCSM) variants in combination with realistic hyperon-nucleon (YN) interactions from chiral effective field theory (EFT) \cite{Wirth2014abinitio,Wirth2018hypernuclear,Wirth2018light,Le2020jacobi}. 
While first applications of those chiral interactions in hypernuclear structure calculations yield promising results, some deficits, such as the systematic overbinding of the hyperon, have become apparent \cite{Wirth2018diss}.
In the strive for a precise and accurate description of hypernuclei, these deficits need to be addressed from multiple directions regarding the underlying interactions as well as the employed many-body methods.
In this work we discuss both of these aspects, in particular, the poorly constrained YN interactions and a systematic extrapolation and uncertainty quantification for hypernuclear observables from calculations in finite model spaces.

Our main focus is on the YN interaction and how the low-energy constants (LECs) can be constrained. 
So far, a YN interaction from chiral EFT has been derived up to next-to-next-to-leading order (N$^2$LO) \cite{Polinder2006hyperon,Haidenbauer2013hyperon,Haidenbauer2020hyperon,Haidenbauer2023hyperon}.
A full description of an initial YNN interaction from chiral EFT, which would occur from N$^2$LO on, is not available for now \cite{Petschauer2016leading}, though, the addition of such three-body forces is generally considered to be important for the accurate description ranging from strange few-body systems to neutron stars \cite{Lonardoni2013effects,Wirth2016induced}. 
Looking at the individual chiral orders of the YN interaction one finds 5 LECs at LO and 23 LECs at NLO and N$^2$LO. 
Analogously to the nucleonic sector, these LECs need to be fitted to experimental data, typically two-baryon scattering data.
However, due to the short-lived nature of hyperons there is very little experimental data available for now. 
The existing determinations of the LECs in the chiral YN interaction use the 35 YN scattering data points and the hyperon separation energy $B_\Lambda$ of \lelem{H}{3}.
While it is nearly impossible to constrain 23 free parameters on 36 data points without imposing additional symmetries, as done in \cite{Haidenbauer2013hyperon,Haidenbauer2023hyperon}, even fixing only 5 LECs on such little data is a challenging task given the poor quality of the data, resulting in a rather ill-constrained interaction.
This naturally raises the question of whether one can exploit other, more precise experimental data to determine the LECs. In recent generations of nucleonic interactions, many-body observables have been included in the determination of the LECs in addition to the previously used scattering data, resulting in a considerably improved description of both ground-state energies and radii in medium-mass nuclei \cite{Ekstrom2015accurate,Jiang2020accurate,Huether2020family}.
Following the same lines we employ experimental data for p-shell hypernuclei as additional constraints on the YN interaction, thus, constructing an interaction that is optimized for hypernuclear structure calculations. 
In order to keep the number of LECs low, we concentrate our investigations on the LO interaction.
We first perform calculations for the hypernucleus \lelem{Li}{7} to study the sensitivities of hyperon separation energies as well as excitation energies to the individual LECs, which allows us to further reduce the active degrees of freedom. 
We then adjust the remaining LECs to optimally describe a selected set of experimentally well-known hypernuclear observables.

Using many-body calculations for the determination of LECs requires a careful assessment of the model-space convergence and control of the many-body uncertainties. In the NCSM there is a single control parameter $N_{\max}$ that controls the model-space dimension and, thus, the convergence and uncertainties. One could resort to classical extrapolation schemes based on exponential modeling \cite{Maris2009abinitio,Bogner2008convergence,Coon2012convergence,Gazda2022nuclear} 
to extract converged observables and associated uncertainties. 
However, novel machine learning tools \cite{Negoita2019deep,Jiang2019extrapolation,Vidana2023machine,Knoell2023machine,Wolfgruber2023prediction} have proved very valuable in this context. 
We adapt the neural network approach presented in \cite{Knoell2023machine,Wolfgruber2023prediction} and show its capabilities for predicting converged hyperon separation energies along with statistically meaningful uncertainty estimates.
While the many-body uncertainties are immediately relevant for the LEC optimization, we should also address the uncertainties resulting from the truncation of the chiral expansion for the interaction. The quantification of interaction uncertainties typically utilizes Bayesian methods \cite{Epelbaum2015improved,Binder2016few,Melendez2017bayesian,Binder2018few,Epelbaum2019few,Melendez2019quantifying} exploiting the order-by-order behaviour of the observable and are, therefore, not suited to estimate an error for a LO interaction.
One could alternatively resort to an uncertainty estimation obtained through a variation of the underlying nucleonic interactions as discussed in \cite{Gazda2022nuclear}. 
However, this still neglects the truncation of the YN interaction, which is the largest source of interaction uncertainties.
In the present paper we, therefore, limit our discussion to many-body uncertainties as we focus on an optimized description of hypernuclei for practical applications based only on the LO contributions of the YN interaction.

We will apply the optimized YN interaction for predictions of ground state properties and excitation spectra of hypernuclei throughout the \lelem{He}{} isotopic chain.

\paragraph{The Hypernuclear IT-NCSM}

Our many-body method of choice is the importance truncated no-core shell model (IT-NCSM) \cite{Navratil2009recent,Roth2009importance,Barrett2013abinitio}.
The stationary Schrödinger equation is cast into a matrix eigenvalue problem by expanding the states in a set of Slater determinants $\qty{\ket{\phi_i}}$
\begin{align}
    \sum_j\mel*{\phi_i}{H}{\phi_j}\braket*{\phi_j}{\psi_n} = E_n\braket{\phi_i}{\psi_n} \quad \forall i,
\end{align}
with Hamiltonian $H$, energy eigenvalues $E_n$, and corresponding eigenstates $\ket{\psi_n}$.
The Slater determinants are constructed from single-particle states in the harmonic oscillator (HO) basis for a given frequency $\hbar\Omega$.
For non-strange nuclei, this single-particle basis is limited to protons and neutrons.
The extension to hypernuclei \cite{Wirth2014abinitio,Wirth2018hypernuclear} is conceptually straight forward by introducing strangeness $\curly{S}$ as a quantum number.
A single-particle state in the HO basis is then given by
\begin{align}
    \ket*{n(ls)jm_j,\curly{S}\,tm_t}.
\end{align}
If we limit the strangeness to $\curly{S}\in\qty{0,-1}$ the baryonic constituents of the basis now include $p,n,\Lambda,\Sigma^{-},\Sigma^{0}$ and $\Sigma^{+}$.
In order to make this eigenvalue problem computationally tractable, the model space is truncated with respect to the number of HO excitation quanta $N_\mathrm{max}$, which controls the basis dimensions and the convergence behaviour.
In addition, an adaptive importance truncation build on a perturbative importance measure can be used to extend the computational range to larger $N_\mathrm{max}$ \cite{Roth2009importance}, which is important since for given total baryon number $A$ and $N_{\max}$ the model-space dimensions for hypernuclei are significantly larger than for nuclei without strangeness.

The hypernuclear Hamiltonian, required as input for the NCSM, consists of a term for the kinetic energy and a nucleon-nucleon (NN) and three-nucleon (3N) part accompanied by a YN interaction.
We further employ a similarity renormalization group (SRG) transformation of the Hamiltonian in order to accelerate the convergence of the NCSM calculation \cite{Roth2010nuclear,Furnstahl2013new,Roth2014evolved,Wirth2019similarity}.
This unitary transformation induces additional terms up to the $A$-body level. 
Here, all the induced 3N and YNN forces are taken into account explicitly, while higher particle ranks are being neglected.

All calculations in this work are performed with the non-local NN+3N interaction from $\chi$EFT at N$^3$LO with cutoff $\Lambda=500$~MeV ($\mathrm{NN}_\mathrm{EMN}+\mathrm{3N}_\text{H}$) discussed in \cite{Huether2020family}.
The starting point for our optimization is the aforementioned chiral YN interaction at LO with cutoff $\Lambda_\mathrm{YN}=700$~MeV ($\mathrm{YN}_\mathrm{P}$) presented in \cite{Polinder2006hyperon}.
Both, NN+3N and YN interactions are consistently SRG evolved to flow parameter $\alpha=0.08$~fm$^4$.

\paragraph{Uncertainty Quantification}

In order to construct a meaningful procedure for the determination of the LECs one needs to address model-space extrapolations and uncertainty quantification.
As already mentioned, we will not attempt to construct an uncertainty estimate for the truncation of the chiral expansion, since we are limited to the YN interaction at LO.
However, we can address the uncertainties coming from the incomplete convergence of the IT-NCSM calculations with respect to model-space size, which arguably become significant in larger p-shell nuclei due to the factorial growth of the model-space dimensions with particle number $A$.

In this work we adapt the machine learning tool presented in \cite{Knoell2023machine,Wolfgruber2023prediction} and show its predictive capabilities for hypernuclei. 
It uses artificial neural networks (ANNs) in order to predict converged observables in the infinite Hilbert space from the convergence patterns in small model spaces.
The ANNs are designed to take three sequences of (IT-)NCSM calculations for different HO frequencies $\hbar\Omega$, each consisting of results for four consecutive model-space sizes $N_\mathrm{max}$.
These networks are then trained on a huge set of fully converged NCSM results for few-body systems up to $A=4$ with different chiral NN+3N interactions and SRG flow parameters.
This way the ANNs learn various different convergence patterns resulting in a universality that enables the application to arbitrary p-shell nuclei, interactions and states.
This universality is also what allows the direct transfer of the ANNs to hypernuclei.
The convergence patterns of ground-state energies in hypernuclei are very much alike the ones in regular nuclei as they are dominated by the same nucleonic interactions.
Due to their interpolation capabilities, ANNs should also be able to account for slight deviations in the convergence patterns induced by the YN interaction.

\begin{figure}
    \includegraphics[width=\columnwidth]{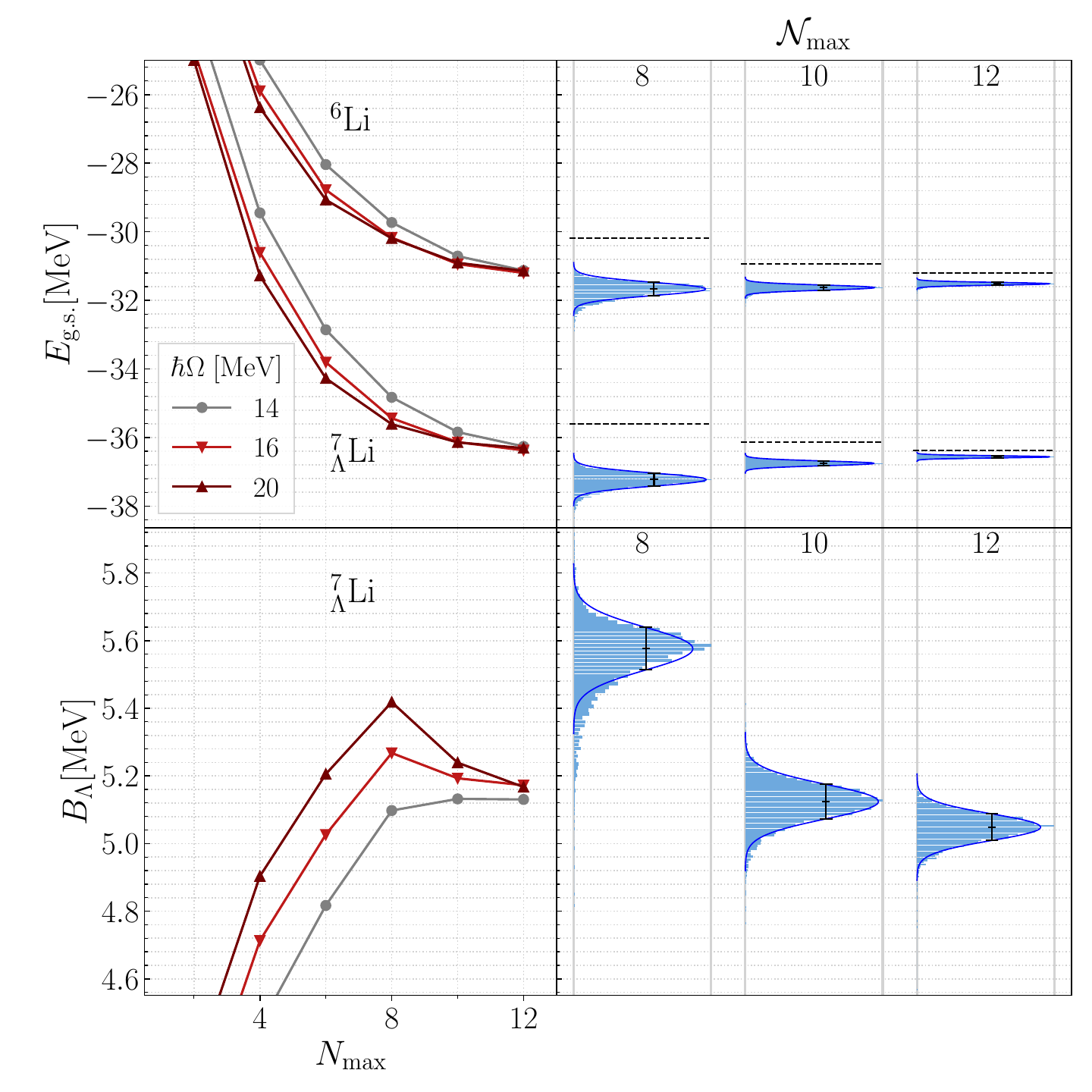}
    \caption{IT-NCSM calculations of ground-state energies for \elem{Li}{6} and \lelem{Li}{7} along with predictions for the converged value from 1000 ANNs at different $\curly{N}_\mathrm{max}$ (upper panel). Hyperon separation energies $B_\Lambda$ for \lelem{Li}{7} along with ANN predictions obtained through subtraction of the predicted ground-state energies of \lelem{Li}{7} and \elem{Li}{6} (lower panel). See text for details. \label{fig:ANN_PRED}}
\end{figure}

The left-hand side of the upper panel in Fig.~\ref{fig:ANN_PRED} shows the results of sequences of (IT-)NCSM calculations for the ground-state energies of \lelem{Li}{7} and its parent nucleus \elem{Li}{6} for different HO frequencies. 
On the right-hand side one finds the corresponding predictions of the converged ground-state energies obtained from 1000 ANNs which were trained
similarly to the ones in~\cite{Knoell2023machine} with the additional step of normalizing the input data to the interval $[0,1]$. 
A detailed discussion of the normalization and its implications on the ANN predictions can be found in~\cite{Wolfgruber2023prediction}.

The distributions of predictions arise from the evaluation of all ANNs with all combinatorically possible samples constructed from the input data at a given $\curly{N}_\mathrm{max}$ which indicates the maximum $N_\mathrm{max}$ in a sample.
The nominal predictions with corresponding uncertainties are then extracted through a fit of a Gaussian to these distributions.
Looking at the results for the ground-state energies we find that the ANNs provide equally plausible predictions for hypernuclei as they do for regular nuclei.

Furthermore, we construct predictions for the hyperon separation energy $B_\Lambda$.
The challenge here is that the convergence behaviour is not constrained by the variational principle as it is the case for ground-state energies, which often leads to a non-monotonic behavior.
To handle this, we construct predictions through a sample-wise subtraction of the predicted ground-state energies for \lelem{Li}{7} and \elem{Li}{6}, which are shown in the lower panel of Fig.~\ref{fig:ANN_PRED}. This procedure is analogous to the prediction of excitation energies discussed in \cite{Wolfgruber2023prediction}.
Looking at the ANN results obtained for $\curly{N}_\mathrm{max}=10,12$ we find very consistent predictions which perfectly agree with what one would expect by looking at the evaluation data. 
The predictions using only the evaluation data up to $\curly{N}_\mathrm{max}=8$ deviate from the ANN predictions obtained by including the larger model spaces. This is an obvious consequence of the anomalous convergence pattern of the separation energies. Even in this case the prediction is plausible, given that the ANNs only see the evaluation data up to $N_\mathrm{max}=8$ that shows a monotonically increasing trend. This observation shows that despite the robustness of the ANNs, data from sufficiently large model spaces is necessary for accurate predictions.

\paragraph{Optimization Procedure}

As already mentioned, the original YN interaction as given in \cite{Polinder2006hyperon} comes with 5 LECs
\begin{align}
\begin{alignedat}{2}
    C^{\Lambda\Lambda}_{^1\mathrm{S}_0}&=-0.0304,\quad
    C^{\Lambda\Lambda}_{^3\mathrm{S}_1}&&=-0.0022,\\
    C^{\Lambda\Sigma}_{^3\mathrm{S}_1}&=0.0035, \quad
    C^{\Sigma\Sigma}_{^1\mathrm{S}_0}&&=-0.0744,\quad
    C^{\Sigma\Sigma}_{^3\mathrm{S}_1}=0.2501
\end{alignedat}
\end{align}
which are associated with particle species and partial waves.
The values given here correspond to cutoff $\Lambda_\mathrm{YN}=700$~MeV for which the interaction performs best regarding hypernuclear structure calculations \cite{Wirth2018light}.
In order to identify the most relevant parameters we first explore how sensitive hypernuclear observables are to changes of individual LECs.
We study these effects on the showcase hypernucleus \lelem{Li}{7} as its hyperon separation energy and the low lying excited states have been measured with good precision.
We do not want to change the LECs to drastically different values, which would result in a bad description of the scattering data they have been fit to initially.
Instead, we change the LECs one at a time on what we consider a natural scale.
In order to identify such a natural scale we look at the average change of the individual LECs for the different cutoffs $\Lambda_\mathrm{YN}=$ 550, 600, 650, and 700~MeV in \cite{Polinder2006hyperon}.
As calculations for hypernuclear structure observables become more accurate for increasing cutoff we further increase the LECs by two times this average change and obtain
\begin{align}
\begin{alignedat}{2}
    C^{\Lambda\Lambda}_{^1\mathrm{S}_0}&=-0.0175,\quad
    C^{\Lambda\Lambda}_{^3\mathrm{S}_1}&&=0.0107,\\
    C^{\Lambda\Sigma}_{^3\mathrm{S}_1}&=0.0060, \quad
    C^{\Sigma\Sigma}_{^1\mathrm{S}_0}&&=-0.0732,\quad
    C^{\Sigma\Sigma}_{^3\mathrm{S}_1}=0.2579.
\end{alignedat}
\label{eq:lecmod}
\end{align}

\begin{figure}
    \hspace{-.4cm}
    \includegraphics[width=1.0\columnwidth]{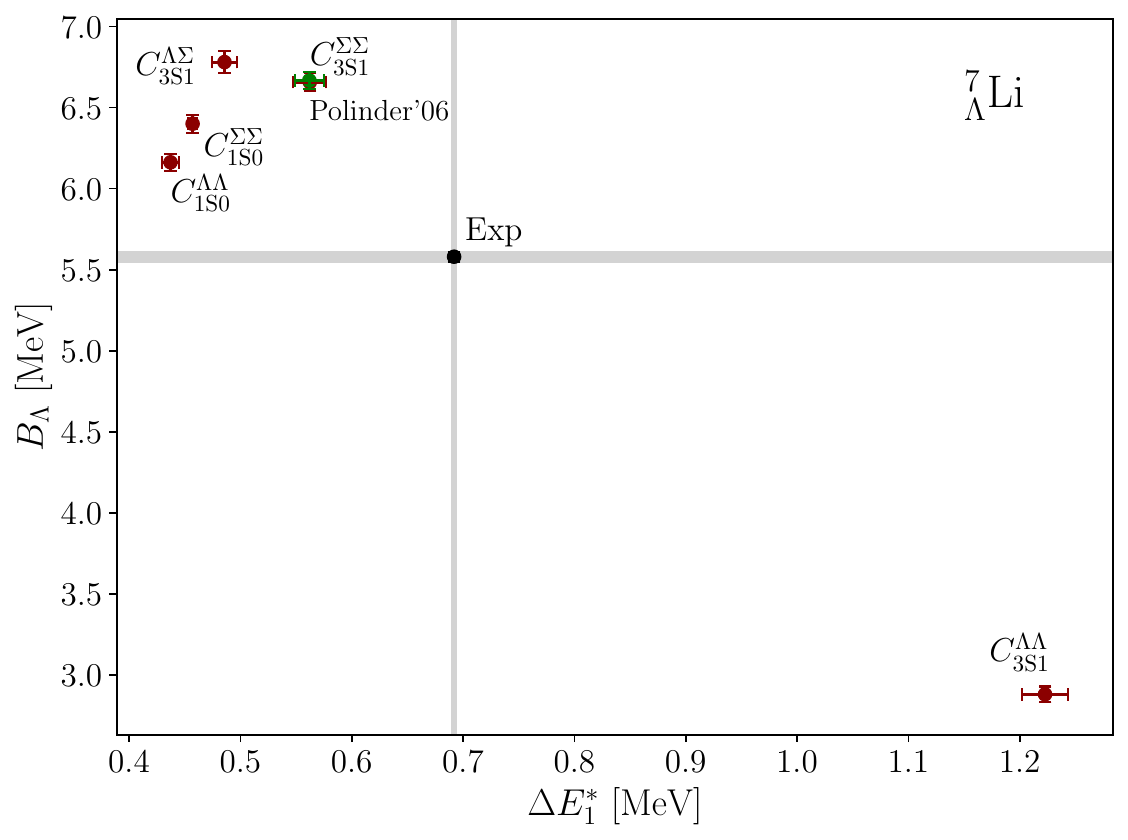}
    \caption{Calculations of hyperon separation energy $B_\Lambda$ and first excitation energy $\Delta E^*_1$ of \lelem{Li}{7} for the original LECs as well as YN interactions with variations of a single LEC indicated by the labels, according to \eqref{eq:lecmod}.  Experimental values are given for comparison \cite{Davis2005years,Hashimoto2006spectroscopy}. \label{fig:SENSITIVITY}}
\end{figure}

Figure~\ref{fig:SENSITIVITY} shows $B_\Lambda$ and the first excitation energy of \lelem{Li}{7} obtained for the YN interactions with a single modified LEC according to \eqref{eq:lecmod}.
One finds that the calculated properties are very sensitive to $C^{\Lambda\Lambda}_{^3\mathrm{S}_1}$ while variations of the LECs associated with $\Sigma$ hyperons show little to no effect.
This is expected since the admixture of $\Sigma$ hyperons to the low-lying states is very small.
We, therefore, limit the parameters for the optimization to $C^{\Lambda\Lambda}_{^1\mathrm{S}_0}$ and $C^{\Lambda\Lambda}_{^3\mathrm{S}_1}$.

For the optimization of the LECs we use a set of experimental data for well-known p-shell hypernuclei, which is predominantly controlled by the YN interaction.
In particular, we use the hyperon separation energies for \lelem{H}{3}, \lelem{He}{5}, \lelem{Li}{7} and \lelem{Be}{9} along with energy differences between spin-orbit partner states in excitation spectra for which we consider the $\frac{1}{2}^+$ and $\frac{3}{2}^+$ states as well as the $\frac{5}{2}^+$ and $\frac{7}{2}^+$ states in \lelem{Li}{7} and the $\frac{3}{2}^+$ and $\frac{5}{2}^+$ states in \lelem{Be}{9}.
Besides being among the best-studied hypernuclei, the experimental values for the hyperon separation energies for the above isotopes scatter significantly and different experiments are not always consistent with each other.
A nice overview of the current experimental situation can be found in \cite{HypernuclearDataBase}.
As a consequence, the selection of experimental values will effect the optimization procedure.
The data chosen here is listed in Table~\ref{tab:summary}.

\begin{figure}
    \hspace{-.4cm}
    \includegraphics[width=1.07\columnwidth]{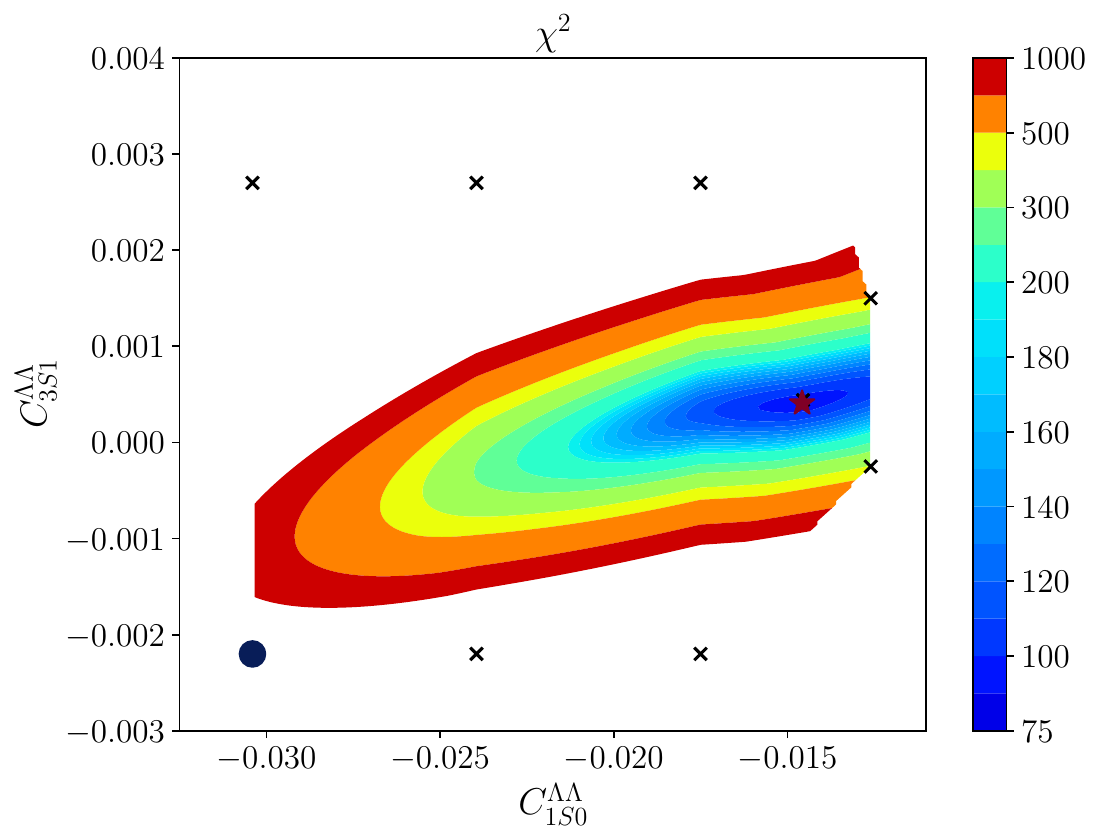}
    \caption{Contour plot of the $\chi^2$ metric in the LEC plane spanned by $C^{\Lambda\Lambda}_{^1\mathrm{S}_0}$ and $C^{\Lambda\Lambda}_{^3\mathrm{S}_1}$. The black crosses ($\cross$) mark the actual calculations, the blue circle (\textcolor{myblue}{$\medbullet$}) indicates the original values of the LECs and the point for the minimal $\chi^2$ is given by the purple star~(\textcolor{mypurple}{$\bigstar$}). The colored area is obtained through interpolation between the black crosses. \label{fig:CONTOURPLOT}}
\end{figure}

The optimization is performed with respect to a $\chi^2$ metric
\begin{align}
    \chi^2 = \sum\frac{(o-o_\mathrm{exp})^2}{\sigma^2_\mathrm{theo}+\sigma^2_\mathrm{exp}}
\end{align}
including experimental and theoretical uncertainties.
Because of the high computational costs for calculating a SRG evolved YN interaction and multiple subsequent IT-NCSM calculations, we construct a grid in the $C^{\Lambda\Lambda}_{^1\mathrm{S}_0}$ - $C^{\Lambda\Lambda}_{^3\mathrm{S}_1}$ plane and interpolate the observables.
The IT-NCSM calculations at the grid points are performed up to $N_\mathrm{max}=14,12,8$ for \lelem{He}{5}, \lelem{Li}{7}, and \lelem{Be}{9}, respectively.
In larger model spaces starting from $N_\mathrm{max}=10,8,6$ we employ the importance truncation.
For the grid point corresponding to the original LECs we perform calculations for three HO frequencies $\hbar\Omega=14,16,20$~MeV to allow for a many-body uncertainty estimation based on the ANN tool discussed earlier.
Note that uncertainties induced by the importance truncation are smaller than the extracted many-body uncertainties and are, therefore, being neglected.
In order to limit the computational effort we assume the many-body uncertainties to be the same at all grid points.
Hence, calculations at the other grid points are only performed for a single HO frequency $\hbar\Omega=16$~MeV.
For the interpolation we use the results obtained in the largest model space accessible for the respective nucleus.
\begin{figure}
    % \hspace{-.1cm}
    \includegraphics[width=1.0\columnwidth]{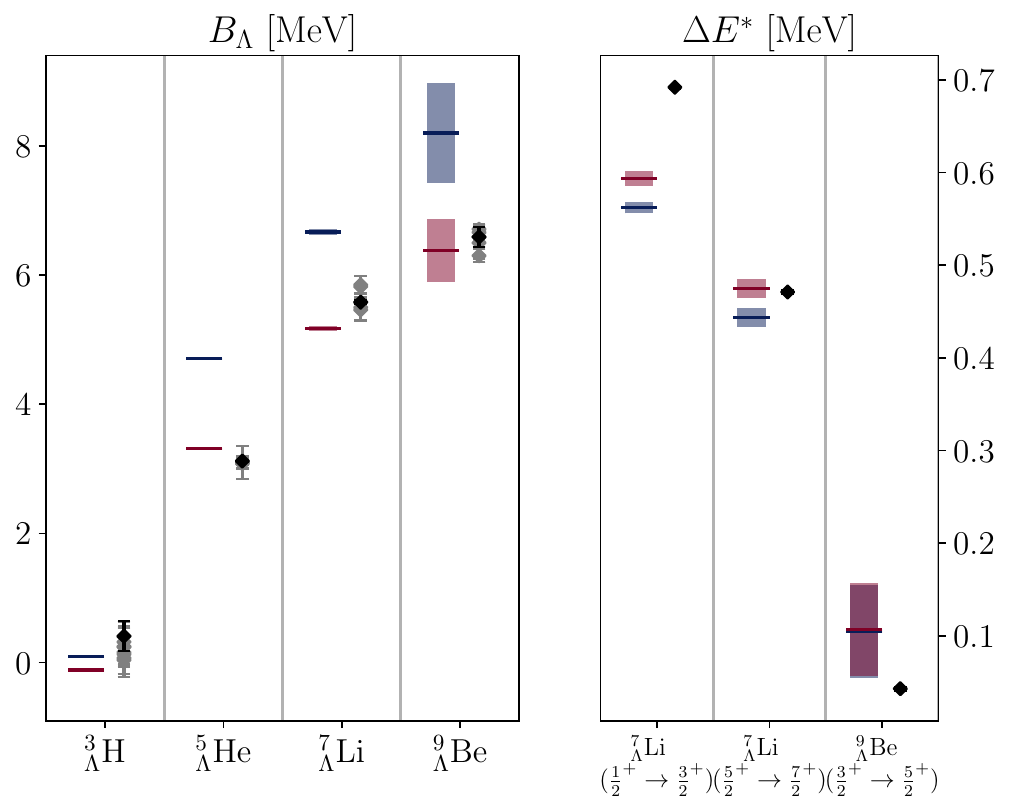}
    \caption{Hyperon separation energies and energy differences of excited states for the p-shell hypernuclei that were included in the optimization process.  Experimental values chosen for the optimization are given in black ($\smallblackdiamond$) and taken from \cite{Davis2005years,Hashimoto2006spectroscopy}, while gray markers (\textcolor{gray}{$\smallblackdiamond$}) indicate other experimental data taken from \cite{HypernuclearDataBase}. Results for the original set of LECs are depicted in blue (\textcolor{myblue}{\textbf{---}}) and results for the optimized interaction in purple (\textcolor{mypurple}{\textbf{---}}). Error bars resemble many-body uncertainties. \label{fig:OBSERVABLES}}
\end{figure}

The combined results of this analysis are shown in Fig. \ref{fig:CONTOURPLOT},
where the $\chi^2$ metric is depicted as a contour plot in dependence of $C^{\Lambda\Lambda}_{^1\mathrm{S}_0}$ and $C^{\Lambda\Lambda}_{^3\mathrm{S}_1}$. The black crosses indicate the grid points the interpolation is constructed on.
In the range which we have considered we find one pronounced minimum at
\begin{align}
    C^{\Lambda\Lambda}_{^1\mathrm{S}_0}=-0.0146,\quad
    C^{\Lambda\Lambda}_{^3\mathrm{S}_1}=0.0004
    \label{eq:optlecs}
\end{align}
with $\chi^2=96$.
Note that the changes compared to the original LECs seem large on a percentage level but are within the same order of magnitude as the previously considered natural range for the respective LEC.
Yet, we want to emphasize that $C^{\Lambda\Lambda}_{^3\mathrm{S}_1}$ changes its sign.
Since the recent experimental value $B_\Lambda=0.41(23)$~MeV for \lelem{H}{3} used in this optimization is disputed, we explore the sensitivity of the optimization result to this particular datum by using $B_\Lambda=0.148(40)$~MeV, which is the currently recommended value from \cite{HypernuclearDataBase}.
This results in a minimum at
\begin{align}
    C^{\Lambda\Lambda}_{^1\mathrm{S}_0}=-0.0153,\quad
    C^{\Lambda\Lambda}_{^3\mathrm{S}_1}=0.0004
\end{align}
with $\chi^2=132.5$.
The modification of $C^{\Lambda\Lambda}_{^1\mathrm{S}_0}$ is much smaller than the change of the LEC compared to the original fit and the effect on the many-body results will be within our uncertainties. 
We, therefore, stick with the values in \eqref{eq:optlecs} for the following investigations.

Generally, the optimized LECs result in a slightly weaker $\Lambda$N interaction.
However, the scattering cross sections obtained with these optimized parameters are compatible with experimental data and deviations from the cross sections obtained with the original LECs are reasonably small \cite{Haid}.

When we take a look at the results of the hypernuclear structure calculations shown in Fig.~\ref{fig:OBSERVABLES} we find that the hyperon separation energies are systematically reduced while the change in the excitation energies is rather small.
Further, the hyperon in the hypertriton becomes slightly unbound, but the description of $B_\Lambda$ in all other hypernuclei is significantly improved. Stated differently, the variation of the dominant LECs at LO alone does not allow for a simultaneous reproduction of the hyperon separation energies of the hypertriton and the heavier p-shell hypernuclei. 
This also holds for the staggering observed for \lelem{He}{5}, being slightly overbound, and \lelem{Li}{7}, which is slightly underbound. These remaining deviations are well within the uncertainties to be expected for a leading order YN interaction and the over-all agreement with experiment is significantly improved compared to the original set of LECs.

\paragraph{Results}

With the optimized interaction and the ANN predictions at hand, we can investigate a broader set of hypernuclei. 
We start with a set of hypernulcei up into the mid-p-shell and then focus on the helium isotopic chains.

\begin{figure}[t!]
    \includegraphics[width=\columnwidth]{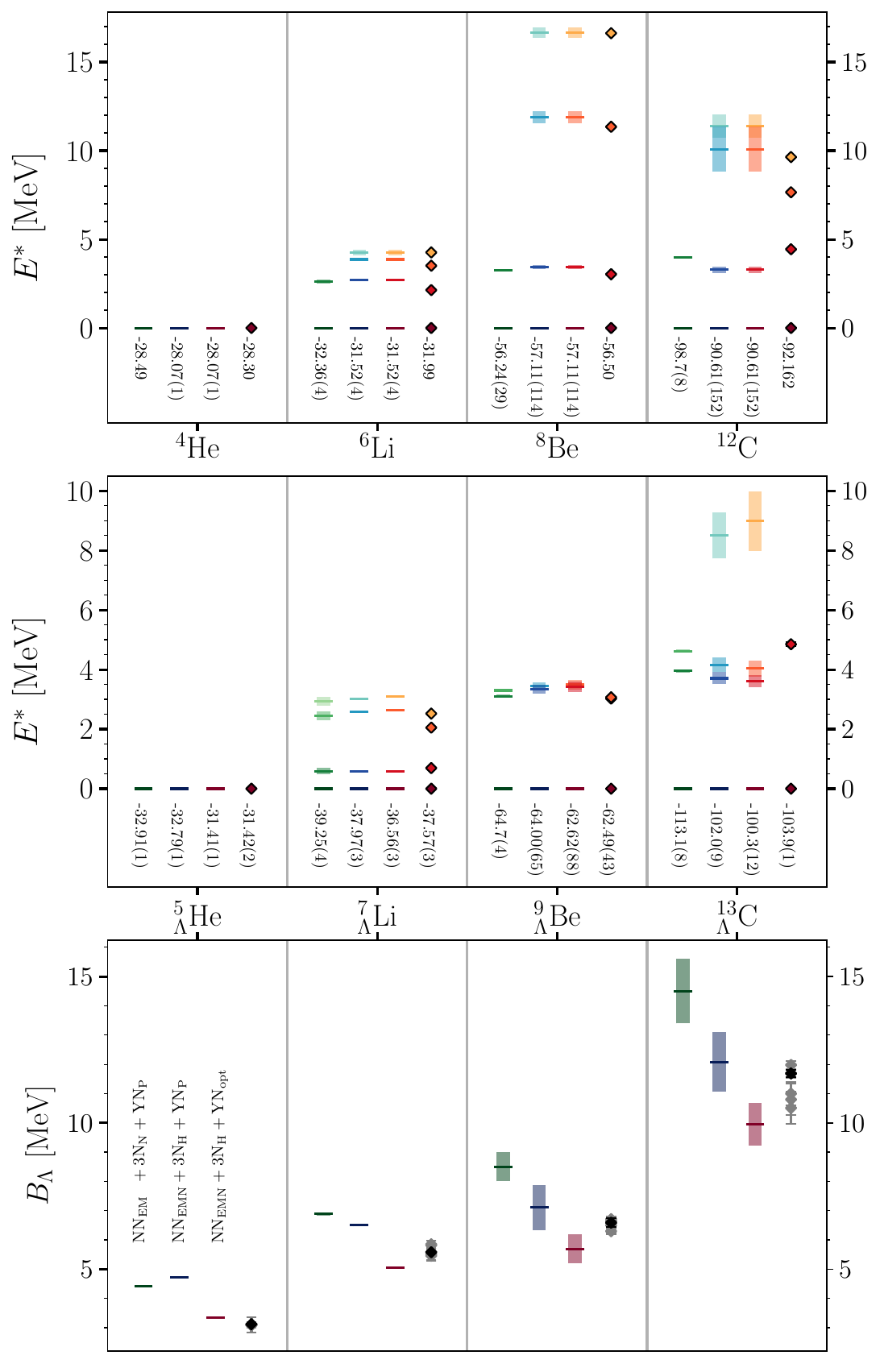}
    \caption{Extrapolated excitation spectra and hyperon separation energies of the low-lying natural-parity states for selected hypernuclei (middle and lower panels) and their parent nuclei (upper panel). Experimental values as given in Table \ref{tab:summary} ($\smallblackdiamond$) are taken from \cite{Davis2005years,Hashimoto2006spectroscopy,Wang2012the,Tilley2002energy,Tilley2004energy,Ajzenberg1990energy}, while the gray markers (\textcolor{gray}{$\smallblackdiamond$}) denote other experimental data taken from \cite{HypernuclearDataBase}. 
    The columns for the individual nuclei indicate, from left to right, results from previous work \cite{Wirth2018diss}, which are calculated with different nucleonic interactions (see text for details), results for the original YN interaction \cite{Polinder2006hyperon} and results obtained with modified LECs.
    Corresponding ground-state energies are denoted below the ground-state markers. Errorbars resemble many-body uncertainties.}
    \label{fig:MISC_RES}
\end{figure}

Figure~\ref{fig:MISC_RES} shows ANN extrapolated ground-state energies and excitation spectra for \lelem{He}{5}, \lelem{Li}{7}, \lelem{Be}{9}, and \lelem{C}{13} and the corresponding parent nucleus, along with the resulting hyperon separation energies.
Calculations for three different combinations of NN+3N and YN interactions are presented in the first three columns of each panel together with the experimental values in the fourth column.
The different interactions include the original YN interaction $\mathrm{YN}_\mathrm{P}$ (middle column) and the modified YN interaction $\mathrm{YN}_\mathrm{opt}$ (right column) in combination with the previously mentioned nucleonic interactions $\mathrm{NN}_\mathrm{EMN}+\mathrm{3N}_\text{H.}$. 
Additionally, results from an earlier work \cite{Wirth2019similarity,Wirth2018light,Wirth2018diss} (left column) are shown for comparison, which were obtained with the same original YN interaction $\mathrm{YN}_\mathrm{P}$ but with a different NN+3N interaction, here denoted as $\mathrm{NN}_\mathrm{EM}+\mathrm{3N}_\text{N}$. This allows for a qualitative assessment of the influence of different nucleonic interactions on hypernuclear observables.
Further, it has to be noted that the uncertainties for the latter are obtained differently from the ANN predictions we employ for the results of this work.

For the discussion of the results, we will first focus on the middle and right columns as they share the same NN+3N interactions. 
Hence, any differences are solely caused by the adjustment of $C^{\Lambda\Lambda}_{^1\mathrm{S}_0}$ and $C^{\Lambda\Lambda}_{^3\mathrm{S}_1}$.
While there is obviously no difference in the excitation spectra of the parent nuclei (upper panel), we also find very little change in the excitation spectra of the hypernuclei (middle panel), which are in overall good agreement with the experimental data.
This consistency in the excitation energies extends to variations of the NN+3N interactions as the results in the left column are very similar to the other results, despite significant differences in some of the ground-state energies.
Moreover, the slight changes in the excitation spectra of the hypernuclei correspond to the shifts of the excitation energies of the parent nuclei and can, therefore, be attributed to the nucleonic interactions.

\begin{figure}[t!]
    \includegraphics[width=\columnwidth]{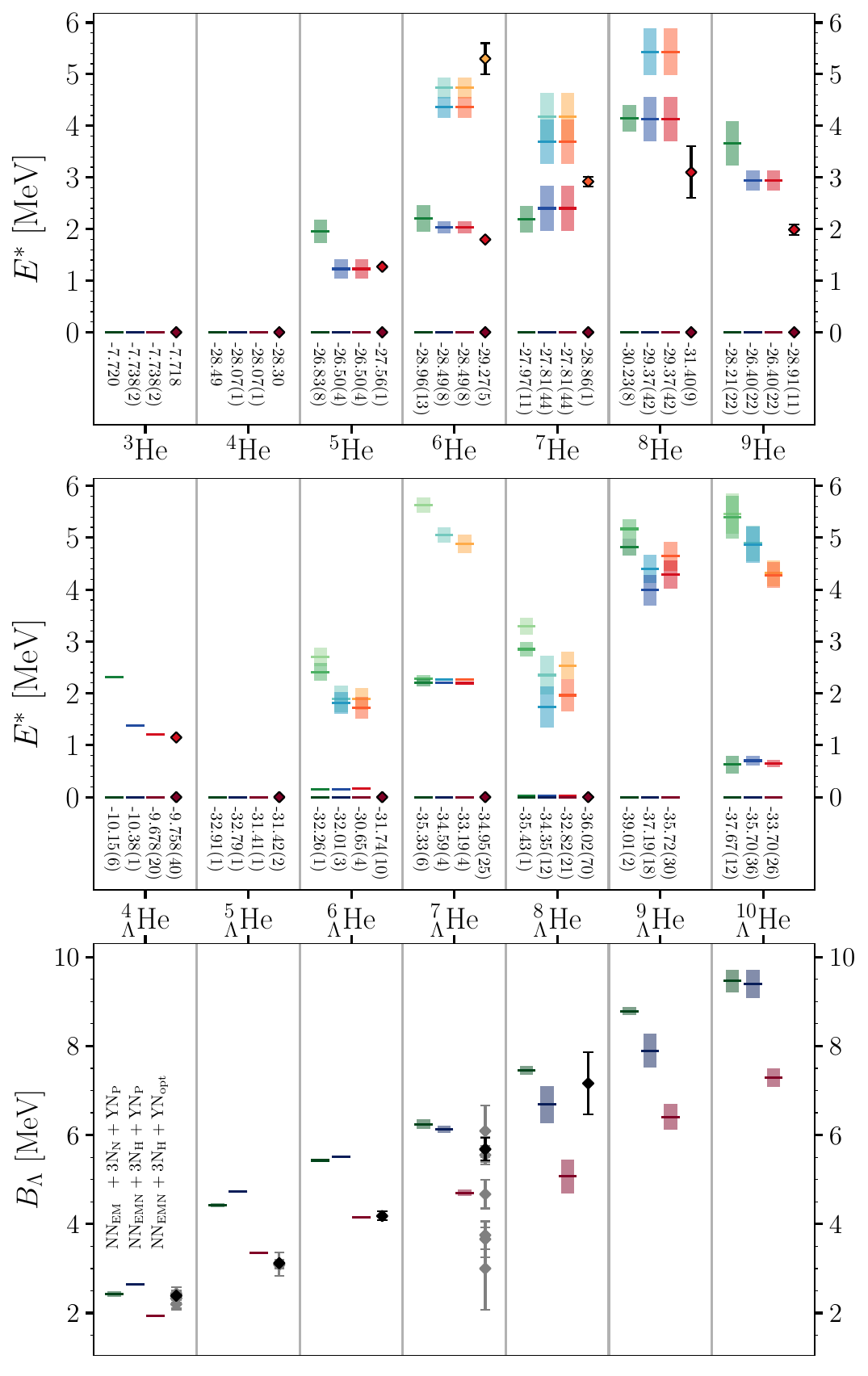}
    \caption{Same as Fig.~\ref{fig:MISC_RES} but for the \lelem{He}{} isotopic chain. Experimental values are taken from \cite{Davis2005years,Hashimoto2006spectroscopy,Nakamura2013observation,Tilley2002energy,Tilley2004energy,Wang2012the,HypernuclearDataBase}.}
    \label{fig:HE_RES}
\end{figure}

When we look at the hyperon separation energies (lower panel) we find that they show a much stronger dependence on the YN interaction.
We, again, see a systematic reduction of $B_\Lambda$ in all hypernuclei for the optimized YN interaction, which is to be expected as \lelem{He}{5}, \lelem{Li}{7} and \lelem{Be}{9} are part of the set of hypernuclei the LECs were optimized on.
Looking at \lelem{C}{13} we see that it is now underbound by the optimized YN interaction, however, the dependency on the nucleonic interaction also becomes much stronger compared to the lighter systems.
This can already be seen for the ground-state energies of \elem{C}{12} which differ by about 10\%. 
Thus, one can hardly make a quantitative statement whether the underbinding is a deficiency of the YN or the NN+3N interaction.

Let us now explore the \lelem{He}{} isotopic chain in order to investigate more nuclei beyond the optimization set.
The extrapolated excitation spectra and hyperon separation energies for the \lelem{He}{} isotopes and their parent nuclei are shown in Fig.~\ref{fig:HE_RES}.

Similar to the previous results we find very little dependence of the excitation spectra on the YN interaction as both the results from the middle and right columns are consistent within their respective uncertainties.
This is in contrast to the results obtained with different NN+3N interactions in the left columns, which deviate much more for most excited states.
The exception is \lelem{He}{10}, where both effects are of similar size.
Contrarily to the hypernuclei discussed before, the changes in the spectra of hypernuclei for the different nucleonic interactions do not always correspond to the shifts of the excited states in the non-strange parent nuclei. New experimental data for spectra of neutron-rich hypernuclei would be needed to assess the quality of the predictions with the optimized YN interaction.

We now turn to the hyperon separation energies in the bottom panel of Fig.~\ref{fig:HE_RES}. We observe that the optimized YN interaction reproduces the experimental separation energies remarkably well for the light isotopes, while for heavier isotopes the separation energies tend to be underestimated.
Simultaneously, the dependence of $B_\Lambda$ on the nucleonic interactions notably increases for neutron-rich isotopes.
Note that the odd Helium isotopes above \elem{He}{4} are unbound and, therefore, difficult to describe within the NCSM.
This discrepancy might translate to the corresponding hypernuclei or be reflected in the hyperon separation energies since they depend on both, parent and hyernucleus.
Similar reasoning applies to \lelem{He}{9} and \lelem{He}{10} which are themselves unbound.

Overall, we find a much better description of hyperon separation energies in \lelem{He}{5,6} and \lelem{Li}{7} with the optimized YN interaction, while the excitation spectra remain mostly unchanged and are in good agreement with experiment.
When turning to the heavier or unbound hypernuclei \lelem{He}{7,8} and \lelem{C}{13} the modified interaction tends to underbind the hyperon, but at the same time dependencies on the nucleonic interactions increase.
A summary of the numerical values for the hyperon separation energies is given in Table~\ref{tab:summary}.

\begin{table}  % --> results with Gen1 ANNs
    \begin{tabular}{ c l l l l}
     \hline\hline
      \multicolumn{2}{c}{NN$_\mathrm{EM}$+3N$_\mathrm{N}$\cite{Entem2003accurate,Navratil2007local}\hspace*{-10pt}} & 
      \multicolumn{2}{c}{NN$_\mathrm{EMN}$+3N$_\text{H}$\cite{Huether2020family}}  \\[-.4em]
      \multicolumn{2}{c}{} & \multicolumn{2}{c}{\hspace{-3pt}-----------------------------} & \\[-.4em]
      \multicolumn{2}{c}{\quad$\mathrm{YN}_\mathrm{P}$} & \multicolumn{1}{c}{\hspace{-10pt}$\mathrm{YN}_\mathrm{P}$} & \multicolumn{1}{c}{$\mathrm{YN}_\mathrm{opt}$} & \multicolumn{1}{c}{Expt.} \\
     \hline\\[-1em]
     \lelem{H}{3} & 0.11(1) & 0.095(5) & -0.092(5) & 0.41(23) \\[.3em]
     \lelem{He}{4} & 2.43(6) & 2.638(11) & 1.940(21) &  2.39(3) \\[.3em]
     \lelem{He}{5} & 4.42(4) & 4.726(8) & 3.348(10) &  3.12(2) \\[.3em]
     \lelem{He}{6} & 5.43(4) & 5.510(33) & 4.151(22) &  4.18(10) \\[.3em]
     \lelem{He}{7} & 6.24(11) & 6.131(79) & 4.701(73) &  5.68(25) \\[.3em]
     \lelem{Li}{7} & 6.89(6) & 6.506(44) & 5.049(39) &  5.58(3) \\[.3em]
     \lelem{He}{8} & 7.46(10) & 6.69(41) & 5.07(38) & 7.16(70) \\[.3em]
     \lelem{He}{9} & 8.78(9) & 7.89(38) & 6.41(29) & -- \\[.3em]
     \lelem{Be}{9} & 8.50(50) & 7.11(77) & 5.69(49) & 6.59(15) \\[.3em]
     \lelem{He}{10} & 9.46(25) & 9.39(32) & 7.29(21) & -- \\[.3em]
     \lelem{C}{13} & 14.5(11) & 12.08(102) & 9.95(72) & 11.69(12)\\ [.3em]
     \hline\hline
    \end{tabular}
    \caption{Predicted hyperon separation energies $B_\Lambda$ for various hypernuclei with many-body uncertainties from ANN predictions. Results are given for the optimized YN interaction ($\mathrm{YN}_\mathrm{opt}$), the unmodified YN interaction ($\mathrm{YN}_\mathrm{P}$) and from previous work \cite{Wirth2018diss}. Experimental data is given for comparison and taken from \cite{Davis2005years,Hashimoto2006spectroscopy,Nakamura2013observation,Tilley2002energy,Tilley2004energy,Wang2012the,Ajzenberg1990energy}.}
    \label{tab:summary}
\end{table}

\paragraph{Conclusions}
In this work we have shown the potential of hypernuclear structure data as additional constraint for the determination of chiral hyperon-nucleon interactions, supplementing the scarce scattering data available. 
With the adjustment of only two LECs in the YN interaction at LO, we have been able to remedy the systematic overestimation of the hyperon separation energy and achieve a precise description of light p-shell hypernuclei in good agreement with experimental data.
Furthermore, we have found a sizable dependence of the hyperon separation energies on the nucleonic interactions, which confirms the findings in \cite{Gazda2022nuclear}.

We have shown that novel machine-learning tools based on ANNs are a valuable extension to the hypernuclear NCSM. They provide robust predictions of converged energies and hyperon separation energies along with meaningful uncertainty estimates for light hypernuclei, while being trained on purely nucleonic systems.
However, a full theoretical uncertainty estimation requires the inclusion of higher chiral orders and a framework to capture the dependency on the nucleonic interactions and on YN interactions consistently.

Ultimately, it will be crucial for realistic YN interactions to include hypernuclear structure data from past and upcoming experiments as additional constraints in the future.
Especially with the increasing number of low-energy constants in higher chiral orders of the YN interaction, this will require a rigorous optimization procedure along with a complete uncertainty quantification.

\paragraph{Acknowledgements}
We thank Johann Haidenbauer for helpful remarks and benchmark calculations of scattering cross-sections.
This work is supported by the Deutsche Forschungsgemeinschaft (DFG, German Research Foundation) through the DFG Sonderforschungsbereich SFB 1245 (Project ID 279384907) and the BMBF through  
Verbundprojekt 05P2021 (ErUM-FSP T07, Contract No. 05P21RDFNB).
Numerical calculations have been performed on the LICHTENBERG II cluster at the computing center of the TU Darmstadt.

\bibliographystyle{elsarticle-num}
\bibliography{bib_hnucl.bib}

\end{document}